\documentstyle[12pt,psfig]{article}
\begin{document}
\title{Multipole Expansion of Bremsstrahlung in Intermediate Energy Heavy Ion
Collisions}

\author{Ulrich Eichmann${}^{a,b}$,Walter Greiner${}^{a}$} 
\date{${}^{a}$Institut f\"ur Theoretische Physik, 
Johann Wolfgang Goethe-Universit\"at, Frankfurt am Main, Germany \\
${}^{b}$Gesellschaft f\"ur Schwerionenforschung mbH, Darmstadt, Germany}
\maketitle

\begin{abstract}
Using a multipole expansion of the radiated field generated by a classical 
electric current, we present a way to interprete the bremsstrahlung
spectra of low energy heavy ion collisions.
We perform the calculation
explicitely for the system ${}^{12}$C+${}^{12}$C at 84AMeV and compare the
result with the experimental data of E.~Grosse et al. 
Using simple model assumptions for the electromagnetic source current we are
able to describe the measured data in terms of coherent photon
emission. In this context, the information contained in the measured data is
discussed. 
\end{abstract}

\section{Introduction}
The angular distribution of bremsstrahlung emitted from accelerated charges
strongly depends upon the structure of the source current. Correlations
between the dynamical properties of the charges show up in the spectrum of
radiated photons. A multipole expansion of the radiated field
\cite{Bied-L,LiA84,Jackson,Bouwkamp}
has the advantage, that special structures like symmetries of the current
enter directly in the coefficients of the expansion. Dominant structures of
the current give dominant terms in the expansion, structures and shapes of
the spectrum provide therefore information about the current. At least one
can conclude from structures in the spectrum, 
that the current is somehow structured as well. Using the multipole
expansion and simple model assumptions for the
electromagnetic source current as well as for the 
geometry and the dynamics of a heavy
ion collision, it is the aim of this paper to show 
that the experimental data of
E.~Grosse et al. 
\cite{EGr85} can be understood as the coherent radiation from a
correlated motion of the nucleons. 
In the following we show how the expansion is obtained. In section
\ref{secstrom} we discuss the current that will be used for the
calculations. With this current the exact and the approximated spectrum are
calculated and compared in section \ref{secspek}. Finally, in section
\ref{datcomp} further assumptions, needed to describe a heavy ion collision,
are made and included in the calculation, which is explicitely performed for
the system ${}^{12}$C+${}^{12}$C at 84AMeV. 

Starting from Maxwell's equations, assuming a periodic time dependence of
the sources, and looking for a solution of Kirchhoff's
differential equation in spherical coordinates one finds a representation of
the electric and the magnetic field in terms of Bessel functions and vector
spherical harmonics. Due to the duality of $\vec E$ and $\vec B$ in
Maxwell's equations one gets two different solutions for $\vec E$ and $\vec
B$ whose linear
combinations are finally the general expressions for the fields.  

More generally one can expand the vector potential, i.e. the wave function
of the photon, in two transversal and
one longitudinal vector field, spanning a 3-dimensional space at each point.
The vector fields are designed in a way that they represent outgoing waves
in the far field. Since a real photon has only two polarizations one only
needs the two transversal fields in the free space outside the sources. 

Thus, the magnetic field reads \cite{Bied-L}:
\begin{equation}
\vec{B}(\vec{r},\omega)=k\sum_{lm}
\left(a^{A(l)\ast}_m(\omega)
\vec{O}^{V(l)}_m(k,\vec{r})
+ a^{V(l)\ast}_m(\omega)\vec{O}^{A(l)}_m(k,\vec{r})\right)
\end{equation}
($k=\mid\vec{k}\mid$),whereas in the far field one gets for the electric field
\begin{equation}
\vec{E}(\vec{r},\omega)=\vec{B}\times \vec{e}(\Omega)\;\;.
\end{equation}
The functions $\vec{O}^{V,A(l)}_m(k,\vec{r})$ are defined by
\begin{equation}\label{os}
\begin{array}{rcl}
\displaystyle \vec{O}^{A(l)}_m(k,\vec{r})&\displaystyle \stackrel{def}{=}&
\displaystyle  \frac{O_l(k,r)}{r}i^l \vec{Y}^{(l,1)l}_m(\Omega)\\
 & & \\
\displaystyle \vec{O}^{V(l)}_m(k,\vec{r})&\displaystyle \stackrel{def}{=}&
\displaystyle  \frac{1}{k}\vec{\nabla}\times\vec{O}^{A(l)}_m(k,\vec{r})
\end{array}
\end{equation}
with $O_l$ the outgoing spherical Bessel function.
The coefficients of the expansion of $\vec{B}$
in terms of the $\vec{O}^{V,A(l)}_m$ read \cite{LiA84}:
\begin{equation}
a^{V,A(l)}_m(\omega)=
\frac{4\pi}{k}\int d^3\vec{r}
\vec{j}(\vec{r},\omega)\vec{F}^{V,A(l)}_m(k,\vec{r})
\end{equation}
and obey the relation
\begin{equation}
\label{stromcc}
a^{V,A(l)\ast}_m(\omega)=(-)^{l+m}a^{V,A(l)}_{-m}(-\omega).
\end{equation}
The $\vec{F}^{V,A(l)}_m(k,\vec{r})$ are defined in the same way as the
$\vec{O}^{V,A(l)}_m$ above with the regular spherical Bessel function
\begin{equation}
F_l(kr)=\frac{kr}{2i^l}\int_{-1}^1e^{ikr\cos\theta}P_l(\cos\theta) 
d\cos\theta\;\;,
\end{equation}
$P_l(\cos\theta)$ being the Legendre polynomials.
The expressions for the coefficients can be simplified by rewriting the
current $\vec{j}(\vec{r},\omega)$ in terms of its Fourier transform and using
Rayleigh's plane wave expansion for the exponential function
\begin{equation}
\begin{array}{rcl}
\vec{j}(\vec{r},\omega) &=&\displaystyle
 \frac{1}{(2\pi)^3}\int d^3\vec{k}\,
e^{i\vec{k}\vec{r}}\,\vec{j}(\vec{k},\omega)\\
 & & \\
 &=&\displaystyle
\frac{2}{(2\pi)^2}\int
k^2dk\sum_{lm}i^{2l}\frac{F_l(kr)}{kr}Y^{(l)\ast}_m(\Omega_r)
\vec{j}^{(l)}_{m}(\omega)
\end{array}
\end{equation}
with 
\begin{equation}
\label{jlm}
\vec{j}(\vec{k},\omega)=\sum_{lm}
i^lY^{(l)\ast}_m(\Omega_k)\vec{j}^{(l)}_m(\omega)\;,\;\;
\vec{j}^{(l)}_m(\omega)=\int d\Omega_k
i^{-l}Y^{(l)}_m(\Omega_k)\vec{j}(\vec{k},\omega)\;\;.
\end{equation}
Further one uses 
$\displaystyle
\vec{Y}^{(l,1)j}_m(\Omega)\equiv
\sum_{m'm''}Y^{(l)}_{m'}(\Omega)\vec{e}{\,}^{(1)}_{m''}
\left(\begin{array}{rl}\left.
\begin{array}{cc}
l&1\\m'&m''\end{array}\right|&\begin{array}{c}j\\m\end{array}
\end{array}\right)$; $\vec{e}{\,}^{(1)}_0=i\vec{e}_z$, $\vec{e}{\,}^{(1)}_{\pm
1}=\frac{\mp i}{\sqrt{2}}\left(\vec{e}_x\pm i\vec{e}_y\right)$
and finally gets
\begin{equation}
\label{coeffs}
\begin{array}{rcl}
a^{V(l)}_m(\omega)&=&(-)^l\left[i^{(l-1)}\sqrt{\frac{l+1}{2l+1}}
\sum\limits_{m'm''}\left(\begin{array}{rl}\left.
\begin{array}{lr}l-1&1\\m'&m''\end{array}\right|&\begin{array}{c}l\\m
\end{array}
\end{array}\right) 
\vec{e}{\,}^{(1)}_{m''}\cdot\vec{j}^{(l-1)}_{m'}(\omega)\right.\\
 & &\left. +i^{(l+1)}\sqrt{\frac{l}{2l+1}}
\sum\limits_{m'm''}\left(\begin{array}{rl}\left.
\begin{array}{lr}l+1&1\\m'&m''\end{array}\right|&\begin{array}{c}l\\m
\end{array}
\end{array}\right) 
\vec{e}{\,}^{(1)}_{m''}\cdot\vec{j}^{(l+1)}_{m'}(\omega)\right]\\
 & & \\
a^{A(l)}_m(\omega)&=&(-)^li^{l}\sum\limits_{m'm''}
\left(\begin{array}{rl}\left.
\begin{array}{lr}l&1\\m'&m''\end{array}\right|&\begin{array}{c}l\\m\end{array}
\end{array}\right) \vec{e}{\,}^{(1)}_{m''}\cdot\vec{j}^{(l)}_{m'}(\omega)\;\;.
\end{array}
\end{equation}
Throughout the calculation we used the orthogonality relations 
\[
\int_0^\infty dr F_l(kr)F_l(k'r)=\frac{\pi}{2}\delta(k-k')
\]
\[
\int d\Omega
Y^{(l)\ast}_m(\Omega)Y^{(l')}_{m'}(\Omega)=\delta_{ll'}\delta_{mm'}\;\;.
\]
The complex conjugate of $\vec{j}^{(l)}_{m}(\omega)$ reads
$\vec{j}^{(l)\ast}_{m}(\omega)=(-)^{m}\vec{j}^{(l)}_{-m}(-\omega)$.

\section{The current}
\label{secstrom}
The classical 4-current of a charged particle moving on a trajectory
$\vec{x}(\tau)$ in proper time can be written as \cite{ItC80}
\begin{equation}                                                           
j^\mu(\vec{x},t)=e\frac{dx^\mu}{dt}\delta^3[\vec{x}-\vec{x}(\tau)]
\mid_{t=x^0(\tau)}=e\int d\tau \frac{dx^\mu}{d\tau}\delta^4[x-x(\tau)]\;\;.
\end{equation}
Rewriting the $\delta$-function by its Fourier integral one finds, that
the current for a colliding particle 
can be understood as the Fourier transform of 
\begin{equation}
\label{fts}
j^\mu(k)=-ie\left(\frac{p^\mu_i}{k\cdot p_i}-\frac{p^\mu_f}{k\cdot
p_f}\right)
\end{equation}
assuming that the collision can be described by a simple kink in the trajectory. 
Coulomb deflection effects, which are important in low-energy collisions,
are neglected here. Generalizing (\ref{fts}) 
to the case of several particles suffering
several collisions we obtain
\begin{equation} \label{gesstrom}
j^\mu(k)=-ie\sum_{i}\sum_j \left(\frac{p^\mu_{i-1j}}{k\cdot p_{i-1j}}-\frac
{p^\mu_{ij}}{k\cdot p_{ij}}\right) e^{ik\cdot x_{ij}}\;\;.
\end{equation}
where the indices count vertices ($i$) and particles ($j$), respectively.
$x_{ij}$ in the equation above accounts for the space-time 
history of the scenario. 
\\

In the following we will make a specific model for the current, considering 
four colliding particles with equal charge and mass. These particles undergo
two subsequent and independent two-body collisions. The first pair interacts
at the space time point $x^\mu=(0,\vec(0))$ and the remaining two collide at
$y^\mu=(y_0,\vec{y})$. For simplicity both collisions are assumed to share the same
reaction plane and absolute value of the deflection angle. 
The Fourier transform of the current in the center of mass frame then reads:
\begin{equation}\label{4strom}
\begin{array}{rcl}
\displaystyle
j_0(\vec k,\omega)&=&
\displaystyle
-\frac{ie}{\omega}\left(T(-a,b)+T(a,-b)-\frac{1}{1+\beta_ix}-
\frac{1}{1-\beta_ix}\right. \\
 & &
\displaystyle
\left.+e^{iky}\left(T(-a,-b)+T(a,b)-\frac{1}{1+\beta_ix}-
\frac{1}{1-\beta_ix}\right)\right)\\
 & & \\
\displaystyle
j_\perp(\vec k,\omega)&=&
\displaystyle
\frac{ie}{\omega}b\left(T(-a,b)-T(a,-b)+e^{iky}\left(-T(-a,-b)+T(a,b)
\right)\right)\\
 & & \\
\displaystyle
j_\|(\vec k,\omega)&=&
\displaystyle
\frac{ie}{\omega}\left(a\left(T(-a,b)-T(a,-b)\right)-
\beta_i\left(\frac{1}{1+\beta_ix}-
\frac{1}{1-\beta_ix}\right)\right.\\
 & &
\displaystyle
\left.+e^{iky}\left(a\left(T(-a,-b)-T(a,b)\right)-\beta_i\left(
\frac{1}{1+\beta_ix}-
\frac{1}{1-\beta_ix}\right)\right)\right)
\end{array}
\end{equation}
with
\[
T(a,b)=\frac{1}{1-ax+b\sqrt{1-x^2}\cos\varphi_k}
\]
$a=\beta_\|$, $b=\beta_\perp$ and $x=\cos\theta_k$.
For sake of simplicity we only consider a temporal distance $\Delta t=y_0$
between the second collision and the first collision. 
The current (\ref{4strom}) has therefore been constructed in a symmetric
manner. For $\Delta t=0$ it is not only
invariant under a parity transformation but also mirror symmetric with
respect to axes both parallel and transverse to the beam direction. These
symmetries are broken with finite $\Delta t$.

\section{The spectrum}
\label{secspek}
The spectrum of bremsstrahlung photons radiated by charged particles
is given by the simple expression
\cite{ItC80}
\begin{equation}\label{spektrum}
\frac{dN^\gamma}{d\tilde{k}}=-\mid j^\mu(k)\mid^2
\end{equation}
with $d\tilde{k}=\frac{1}{(2\pi)^3}\frac{d^3
k}{2k^0}$.
The angular dependence of the spectrum is determined by the scalar
products in the denominators of (\ref{gesstrom}) $p^\mu k_\mu =
p_0k_0-|\vec{p}||\vec{k}|\vec{e}(\Omega_p)\cdot\vec{e}(\Omega_k)$,
$\vec{e}(\Omega_p)\cdot\vec{e}(\Omega_k)=\cos\theta_p\cos\theta_k+
\sin\theta_p\sin\theta_k\cos(\varphi_k-\varphi_p)$.
\\

The spectrum of radiated energy of the process (\ref{4strom}) therefore reads
($I=\omega N$)
\begin{equation}\label{spektrumII}
\frac{dI}{d\omega d\Omega_k}=-\frac{\omega^2}{2(2\pi)^3}\mid
j^\mu(k)\mid^2\;\;.
\end{equation}
Inserting the current (\ref{4strom}) and averaging 
over the azimuthal angle $\varphi_k$, leads to  
\begin{eqnarray}
\label{specex}
\frac{dI}{d\omega d\Omega_k}&=&-\frac{\alpha}{2\pi^2}
\Bigg[\frac{8\cos^2(\frac{\omega \Delta
t}{2})}{(1-\beta_i^2x^2)}
\left(
\frac{(1-\beta_i^4x^2)}{(1-\beta_i^2x^2)}-(1+a\beta_i^2x)Q(a,b,x)+
(1-a\beta_i^2x)Q(-a,b,x)\right)\nonumber \\
 & &+\frac{\cos(\omega \Delta t)}{ax}\left(
\frac{(-1-a^2+b^2-2ax(a^2-b^2))}{1+ax}Q(a,b,x)+\right.\nonumber\\
 & &\hphantom{+\frac{\cos(\omega \Delta t)}{ax}}\left.
\frac{(1+a^2-b^2-2ax(a^2-b^2))}{1-ax}Q(-a,b,x)\right)\nonumber\\
 & &+(1-\beta_f^2)((1+ax)Q(a,b,x)^3+(1-ax)Q(-a,b,x)^3)\nonumber\\
 & &+(1+\beta_f^2)(Q(a,b,x)+Q(-a,b,x))\Bigg]
\end{eqnarray}
with the abbreviation $Q(a,b,x)=\sqrt{(1+ax)^2-b^2(1-x^2)}$.\\

The photon spectrum is given by the Poynting vector,
i.e.
\begin{eqnarray}
\frac{dI}{d\omega d\Omega_k}&=& \frac{1}{2\pi}
\mid r \vec{B}(\omega)\mid^2\nonumber\\
 &=&\frac{r^2k^2}{2\pi}\left| \sum_{lm}
\left(a^{A(l)\ast}_m(\omega)
\vec{O}^{V(l)}_m(k,\vec{r})
+ a^{V(l)\ast}_m(\omega)
\vec{O}^{A(l)}_m(k,\vec{r})\right)\right| ^2
\end{eqnarray}
After changing the order of summation
$\sum\limits_{lm}=\sum\limits_{l=0}^{\infty} \sum\limits_{m=-l}^{l}
\to \sum\limits_{m=-\infty}^{\infty} \sum\limits_{l=0}^{|m|}$ and averaging
over the azimuthal angle $\varphi$, i.e. averaging over all possible 
reaction planes, terms containing the product of two $\vec{O}^{V,A(l)}_m$ 
with different $m$ drop out,
which yields:
\begin{equation}
\label{multspec}
\frac{dI}{d\omega d\Omega_k}=\frac{r^2k^2}{2\pi}\sum_{m} \left|\sum_{l}
...\right|^2\;\;.
\end{equation}
We will describe the spectrum with the nonvanishing terms of lowest order in
$l$. In a similar way the photon emission was calculated for nuclear
collisions below the Coulomb barrier \cite{JR}.\\

The occuring products of the vector fields $\vec{O}^{A,V(l)}_m(k,\vec{r})$
determine the angular dependence of the spectrum. Due to the averaging over
$\varphi$ these products are real and depend only upon $\cos\theta_k$.\\
We use the far field limit for the functions $\vec{O}$:
\begin{eqnarray}
\vec{O}^{A(l)}_m(k,\vec{r})&\approx&
\frac{e^{ikr}}{r}\vec{Y}^{(l,1)l}_m(\Omega_k)\nonumber\\
\vec{O}^{V(l)}_m(k,\vec{r})&\approx&
-\frac{e^{ikr}}{r}\left\{\sqrt{\frac{l+1}{2l+1}}\vec{Y}^{(l-1,1)l}_m(\Omega_k)
+\sqrt{\frac{l}{2l+1}}\vec{Y}^{(l+1,1)l}_m(\Omega_k)\right\}
\end{eqnarray}
Both, the product $\vec{O}^{V(l)\ast}\vec{O}^{V(l)}$ and 
$\vec{O}^{A(l)\ast}\vec{O}^{A(l)}$,
respectively, yield the well known angular functions \cite{Jackson}
\begin{equation}
\begin{array}{ll}
\displaystyle
\frac{3}{r^28\pi}\sin^2\theta_k&(l=1,m=0)\\
 & \\
\displaystyle
\frac{3}{r^216\pi}\left(1+\cos^2\theta_k\right)&(l=1,m=\pm 1)\\
 & \\
\displaystyle
\frac{15}{r^28\pi}\cos^2\theta_k\sin^2\theta_k&(l=2,m=0)\\
 & \\
\displaystyle
\frac{5}{r^216\pi}\left(1-3\cos^2\theta_k +4\cos^4\theta_k\right)
&(l=2,m=\pm 1)\\
 & \\
\displaystyle
\frac{5}{r^216\pi}\left(1-\cos^4\theta_k\right)&(l=2,m=\pm 2)
\end{array}
\end{equation}
For equal $l$ the interference terms
are
asymmetric functions of $\cos\theta_k$. The evaluation yields for $l=2$
\begin{equation}
\label{inter22}
\begin{array}{l}
\displaystyle
\vec{O}^{V(l)\ast}_m(|k|,\vec{r})\cdot\vec{O}^{A(l)}_m(|k|,\vec{r})=\\
\\
\displaystyle
\frac{(-)^{1+m}}{r^2 4\pi}\sqrt{\frac{5}{2}}
\left(P_1(\cos\theta_k)
\left(\begin{array}{rl}\left.
\begin{array}{lr}2&2\\m&-m\end{array}\right|&\begin{array}{c}1
\\0\end{array}\end{array}\right)
+2P_3(\cos\theta_k)
\left(\begin{array}{rl}\left.
\begin{array}{lr}2&2\\m&-m\end{array}\right|&\begin{array}{c}3
\\0\end{array}\end{array}\right)\right)
\end{array}
\end{equation}
and for $l=1$ (only $m=\pm 1$)
\begin{equation}
\label{inter11}
\vec{O}^{V(l)\ast}_{\pm 1}(|k|,\vec{r})\cdot\vec{O}^{A(l)}_{\pm 1}(|k|,\vec{r})=
\mp \frac{3}{r^2 8\pi}P_1(\cos\theta_k)
\end{equation}
For symmetric charge distributions, i.e. currents with even parity, the
coefficients of these terms vanish. \\

The interference of dipole ($l=1$) and quadrupole ($l=2$) contributes with
two different kinds of terms: both, the interference of magnetic and
electric components of $\vec{B}$ and the interference of magnetic and
magnetic as well as electric and electric components, respectively.
The latter is asymmetric in $\cos\theta_k$ \cite{JR}
\begin{equation}
\label{inter12}
\begin{array}{l}
\displaystyle
\vec{O}^{V(1)\ast}_m(|k|,\vec{r})\cdot\vec{O}^{A(2)}_m(|k|,\vec{r})\\
\\
\displaystyle
=\vec{O}^{V(2)\ast}_m(|k|,\vec{r})\cdot\vec{O}^{A(1)}_m(|k|,\vec{r})\\
\\
\displaystyle
=\frac{(-)^{1+m}}{r^2 4\pi}\sqrt{3}\left(\sqrt{\frac{3}{2}}P_1(\cos\theta_k)
\left(\begin{array}{rl}\left.
\begin{array}{lr}1&2\\m&-m\end{array}\right|&\begin{array}{c}1
\\0\end{array}\end{array}\right)
+P_3(\cos\theta_k)
\left(\begin{array}{rl}\left.
\begin{array}{lr}1&2\\m&-m\end{array}\right|&\begin{array}{c}3
\\0\end{array}\end{array}\right)\right)
\end{array}
\end{equation}
Hence, for a current with even parity the interference reduces to the
product
\begin{equation}
a^{A(1)}_{\pm 1}(\omega)
\vec{O}^{V(1)\ast}_{\pm 1}(|k|,\vec{r})\cdot
a^{V(2)\ast}_{\pm 1}(\omega)\vec{O}^{A(2)}_{\pm
1}(|k|,\vec{r})=
\pm \frac{1}{r^2 4\pi}\sqrt{\frac{15}{4}}P_2(\cos\theta_k)
\end{equation}

\begin{figure}[hbpt]
\centerline{\psfig{figure=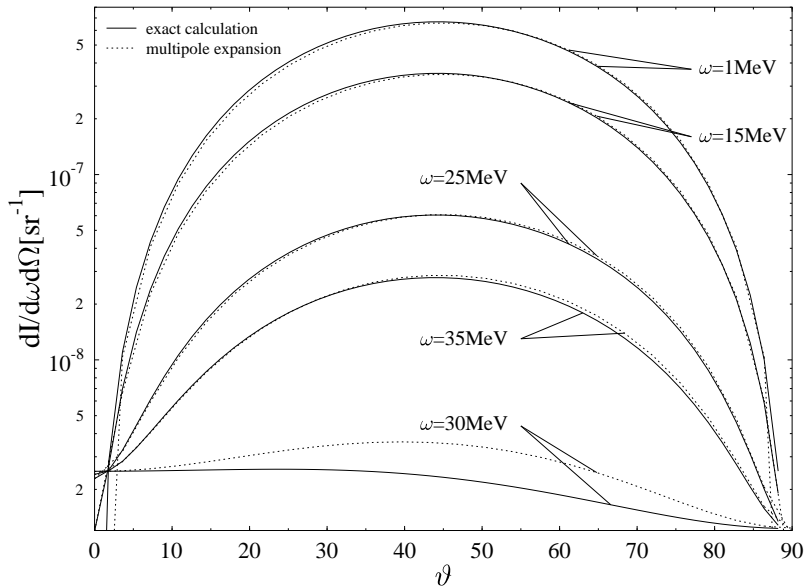,height=10cm}}
\caption{\label{speccomp}Comparision of the exact 
bremsstrahlung spectrum and the spectrum calculated by the first terms of
the multipole expansion for a scattering angle $\theta_p=3.6^\circ$, the 
initial rapidity $y=0.2$ and the final velocity $\beta_f=0.5\beta_i$,
for $\Delta t$ we chose the large value of 20fm/c. 
The spectra are symmetric with respect to
$\theta_k=\pi /2$. 
The transverse scattering generates finite radiation at $\theta_k=\pi /2$
even for soft photons ($l=2$, $m=2$ moment), the temporal phase between the
two pairs of interaction causes the oscillation of the different terms in
the expansion, i.e. the increase of the $l=2$, $m=1$ moment and therefore
photon radiation in beam direction.}
\end{figure}

Here and in eq.(\ref{inter11}) we used that we restricted the motion
of the particles to a plane. The scalar product of
$\vec{j}^{(l)}_m(\omega)$ with $\vec{e}{\,}^{(1)}_{\pm 1}$ in
eq.(\ref{coeffs}) becomes a scalar product with $\vec{e}_x$ or $\vec{e}_y$.
Choosing, without loss of generality, the $x-z$-plane as scattering plane, 
the coefficients
$a^{A(l)}_0(\omega)$ vanish. For symmetric collisions the coefficients
$a^{V(1)}_m(\omega)$ and $a^{A(2)}_m(\omega)$ vanish, too. The possibility
to substitute $\vec{e}{\,}^{(1)}_{\pm 1}$ by $\sqrt{\frac{1}{2}}\vec{e}_x$
further simplifies the calculation since now complex conjugation of the
coefficients $a^{A,V(l)}_m(\omega)$ and the
switching of the sign of the index $m$ decouple (see eq. (\ref{stromcc})).

A comparision of the exact spectrum (\ref{specex}) and the first terms of
the multipole expansion (\ref{multspec}) in Figure \ref{speccomp} 
shows the quality of the
restriction to terms of lowest order in $l$. 
This restriction is valid as long as the initial and final rapidities can
be described well enough by 
\begin{equation}
\label{rapiex}
y\approx \beta +\frac{\beta^3}{3}
\end{equation}
For higher energies, terms of higher order in $l$ have to be taken into
account, i.e. the bremsstrahlung spectrum of ultrarelativistic particles is
composed of infinitely many terms of the multipole expansion and their
interference. 
Although the approximative
character of this treatment is obvious, the multipole expansion offers a
very instructive way to understand the behaviour of
bremsstrahlung emission. 

\section{Comparision with data}
\label{datcomp}
To compare the spectrum obtained with the multipole expansion with
experimental data, one needs to make further assumptions. 
We will perform an analysis of 
the radiation measured in
the system ${}^{12}$C+${}^{12}$C at 84AMeV by Grosse et al \cite{EGr85} who
determied 
the 
photon emission in minimum bias collisions. One therefore has to average the
spectrum over all impact parameters $b$, weighted with the geometric 
cross section of the collision\footnote{One assumes that, as it is valid for
soft photons, the cross section of the radiation process factorizes into the
cross section for the scattering process times the photon spectrum.}:
\begin{equation}
\frac{d\sigma}{d\omega d\Omega}=
\frac{1}{\omega}\int_0^{2\pi}d\varphi \int_0^{bmax}bdb\frac{dI}{d\omega
d\Omega}\;\;\;.
\end{equation}
$b_{max}$ can be chosen as $2R+x$, where $R$ is the radius
of the colliding nuclei and $x$ is the surface thickness of the mass 
distribution.

We treat noncentral heavy ion collisions in the way proposed in
\cite{Westfall}, i.e. the charge $Z(b)$ involved in the collision is
obtained from the geometric overlap integral \cite{Bjorken}
\begin{equation}
V(b)=\int d^3x \,\theta(R^2-x^2-y^2-z^2)\theta(R^2-y^2-(z-b)^2)
\end{equation}
and hence $Z(b)=Z\frac{V(b)}{V}$, $V$ and $Z$ are the volume and the 
charge of the nucleus, respectively.\\
Earlier calculations exhibit the dipole
structure\footnote{According to the remark given in the following of
eq.(\ref{rapiex}) the spectra of high-energy collisions should neither be
called dipolar nor quadrupolar, unless these names refer to the first
nonvanishing electric component of the expansion} of the
photon spectrum, assuming either the validity of incoherent 
summation due to
an uncorrelated motion of the nucleons \cite{Ko} 
or contributions from nucleus nucleus
collisions at large impact parameters whose radiation should possess a
dipole-like structure as well \cite{Vasak86}. Even a mixture of coherent and
incoherent radiation has been suggested for the description of the radiation
\cite{Stahl}. In the following we want to propose as an alternative
mechanism radiation produced by the 
correlated, collective motion
of the nucleons. The current (\ref{4strom}) is considered as a coherent 
elementary process in the heavy ion collision. 
For that reason we constructed the current (\ref{4strom}) in a symmetric
manner since the averaged final state of a symmetric heavy ion collision is
symmetric as well. Eq.~(\ref{4strom}) represents the simplest current which
serves our purposes.
We account for all processes 
which can be described by (\ref{4strom}) considering one to four charged
particles. For collisions involving neutral particles 
the corresponding terms in (\ref{4strom}) are
simply set equal to
zero. We further assume that all nucleons collide at the origin, i.e. the
current simply has to be multiplied by a temporal phase $e^{i\omega t}$ and
integrated over time. The
resulting spectrum of the nucleus nucleus collision is then similar to that
of the nuclei feeling a box-like force $\sim
\theta(T/2-\tau)\theta(\tau +T/2)$ in the time interval $T$, which 
here represents the time of the collision. 
This time dependence of the collision is also obtained,
when one treats the collision of the two nuclei in the frame of a shock wave
model, neglecting the spatial distance of the two outward travelling
shock fronts, i.e. one performs the calculation as if the two shock fronts
were sitting at rest in the origin (with this assumption, radiation is
calculated in \cite{Lippert}).\\
Adopting the shock wave picture, we can estimate the minimal collision time
with a shock velocity $\beta_{sh}$ equal to the speed of light, 
\begin{equation}
\label{tmin}
T_{min}=\frac{D}{\gamma_i(\beta_i+1)}\;\;,
\end{equation}
$D$ is the
diameter of the nuclei.
For noncentral collisions the time is taken to depend linearly on the
diameter at $b/2$.

To obtain the correct behaviour of the spectrum with respect to the photon
energy, one can introduce a phase space correction \cite{Haglin} 
\begin{equation}
\label{pscorr}
\frac{R_2(\tilde{s})}{R_2(s)}=\frac{\lambda^{1/2}(\tilde{s}, m_1^2,
m_2^2)}{\lambda^{1/2}(s,m_1^2,m_2^2)}\frac{s}{\tilde{s}}
\end{equation}
with $\tilde{s} = s - 2\omega \sqrt{s}$ and $\lambda(x,y,z)=(x-y-z)^2-4yz$
the kinematical triangle function; $m_1$ and $m_2$ are the masses of the
colliding particles, respectively.
This way of treating the spectrum of high-energy photons is strictly
justified only for the incoherent summation of the spectrum. Since no adequate
formalism is available at the moment, however, 
we adopt this correction factor
(\ref{pscorr}) also for coherent emission, i.e. we
multiply the resulting spectrum by (\ref{pscorr}). 
To describe the high-energy tail of the spectrum it is unavoidable to
consider either three- and four-nucleon clusters which feed their energy to
the photon production in a collective manner, as proposed in \cite{Shyam}.
We increase the nucleon momentum by the mean 
Fermi-momentum which has to be taken into account since it is of the 
same order of magnitude in the considered system. Consequently, the
available phase space of the photons gets enlarged. \\
In additionally to the decreasing available phase-space, the time dependence of
the collision determines the emission of high energy photons, too. Since the
energy distribution of the spectrum is the square of the 
Fourier-transform of the acting
force, the spectrum of the considered process is poportional to 
\begin{equation}
\frac{dI}{d\omega d\Omega}\propto \frac{1}{\omega^2 T^2}\sin^2(\omega T)
\end{equation}
Since $\lim\limits_{x \to 0} \frac{\sin(x)}{x}=1$ and both, $\omega$ and $T$
enter symmetrically in the spectrum, one obtains the soft photon limes for
either $\omega \to 0$ or $T \to 0$ (for all photon energies). Therefore, for
short collision times one obtains a nearly constant, flat spectrum. In this
way, the
collision time controlls the slope of the spectrum for not too
large photon energies. A lower bound for $T$, however, is given by
(\ref{tmin}) and has the approximate value of 5fm/c.

\begin{figure}[htbp]
\centerline{\psfig{figure=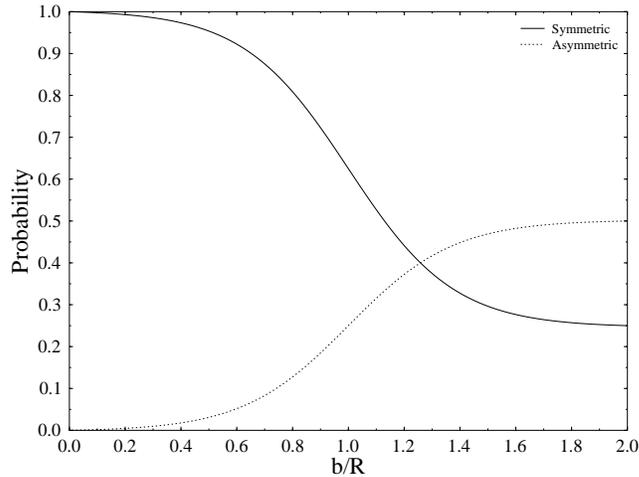,height=8cm}}
\caption{\label{asympara}Parametrization of the probability for asymmetric
and symmetric collisions as a function of the impact parameter. For
convenience, the functions used here are for both cases a hyperbolic
tangent. For symmetric collisions: 
$P_{symm}=\frac{1}{a}\frac{3}{8}{\rm tanh}({\rm
artanh}(a)\cdot ((b/R)-1))+\frac{5}{8}$ and for asymmetric collisions:
$P_{asymm}=\frac{1}{a}\frac{1}{4}{\rm tanh}({\rm
artanh}(a)\cdot ((b/R)-1))+\frac{1}{4}$. $a$ determines the stiffness of the
parametrization and is taken to be $a=0.99$.}
\end{figure}

As mentioned earlier, the presence of 
asymmetric collisions can be expected,which are the consequence of 
fluctuations of the charge distribution in the nuclei. 
The probability for these asymmetries of the charge distribution in the
overlapping volume as a
function of the impact parameter we parametrize in the following way: 
For nearly central
collisions it is zero (the probability for a symmetric collision is 1) and
smoothly drops to 1/2 when the mean charge of one nucleus 
in the overlapping volume becomes $\le
1$. The probability for symmetric collisions approaches 1/4, the remaining
fourth is the probability for symmetric but radiationless collisions, 
i.e. neutron neutron collisions (see Figure \ref{asympara}). 

As possible asymmetric and symmetric
collisions we consider all possible combinations of up to four charges in
(\ref{4strom}) with one exception: instead of the collision of only two 
charged 
particles (one from the projectile and the target)
possessing a temporal distance $\Delta t$ between the scattering
processes, we consider the symmetric case with two particles at each
case, since it is unlikely that all particles scatter in only one
transverse direction. The nonvanishing coefficients are, besides a
factor two which is canceled since only half of the charge enters in the
latter case in the calculation, in both cases the same. In the latter
however only coefficients with even $m$ exist.\\
The "delay"-parameter $\Delta t$ breaks possibly present symmetries of the
scattering process. In this way $\Delta t$ controlls (with respect to
$\omega$) the strength of the contribution of different terms of the
expansion. Processes with a symmetric charge distribution in the overlapping
volume of the colliding nuclei can therefore (for certain photon energies) 
generate a purely dipolar structure in the photon emission. This observation
is similar to the result in \cite{Vasak86}.\\
The asymmetric collisions generate also the interference terms of types
(\ref{inter22})-(\ref{inter12}). However, since the probabilities of the
asymmetries are invariant under a parity transformation, these terms cancel
when averaging over all possible asymmetries. \\
A quantitative evaluation shows that the influences of
different parameters counterbalance to a certain extent. These are 
e.g. the collision time and the cluster size for higher photon energies and 
$\Delta t$ (for intermediate photon energies) or the strength of the stopping (for
soft photons), respectively, and the ratio of symmetric to
asymmetric collisions.

\begin{figure}[hbtp]
\centerline{\psfig{figure=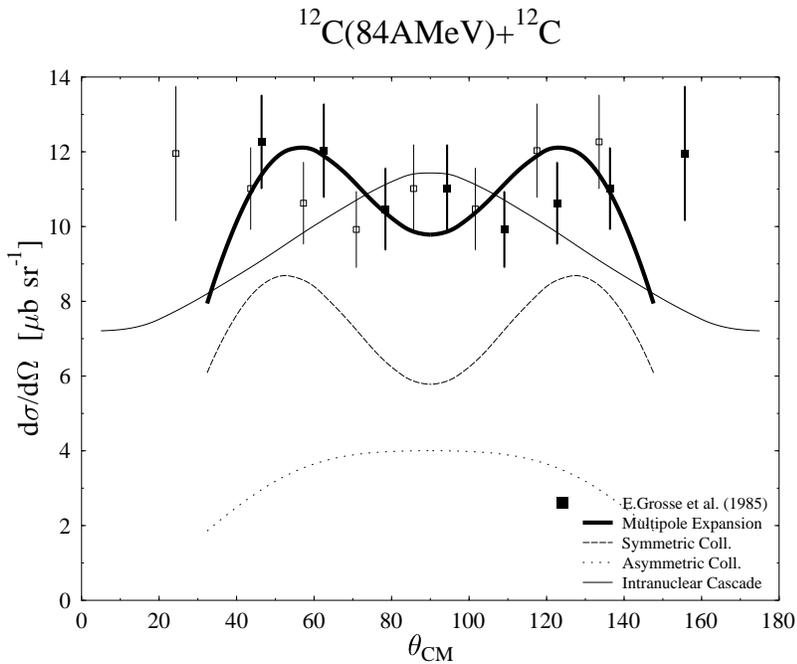,height=10cm}}
\caption{\label{ccspeksum}The bold line shows the spectrum obtained with the
dipole ($l=1$) and quadrupole ($l=2$) terms and their interference.
Contributions of collisions with an assumed asymmetric or symmetric charge
distribution in the nuclear overlap region are shown as well. The
angular distribution of the data 
\protect\cite{EGr85} (symmetrized  with
respect to $\theta =90^\circ$, open squares) can be described quite well.
The thin line represents the result of an intranuclear cascade model
\protect\cite{Ko}. The cross section shown here is integrated over $\omega$
between 50 and 100 MeV. }
\end{figure}

Figure \ref{ccspeksum} shows the spectrum obtained with the discussed
assumptions (full line). The dashed line represents the spectrum calculated
in the frame of the intranuclear cascade model \cite{Ko}. The angular
distribution is not of dipole type and represents the interplay
of the different multipole components and their mutual
interference. Confronting the data with the calculation, the   
measurement seems to indicate an even stronger contribution of the
quadrupole. 
One may expect that the photon angular distribution gets modified when a
larger number of interacting particles is included in the source current
to allow for other
asymmetries with a different ratio of the $l=1$ and $l=2$ terms. With better
knowledge of the charge fluctiations
one can improve the assumed impact parameter dependent ratio of
asymmetric and symmetric collisions. 
E.g., on has to estimate the influence
of the repulsive Coulomb forces on the protons in the nuclei.

\begin{figure}[hbtp]
\centerline{\psfig{figure=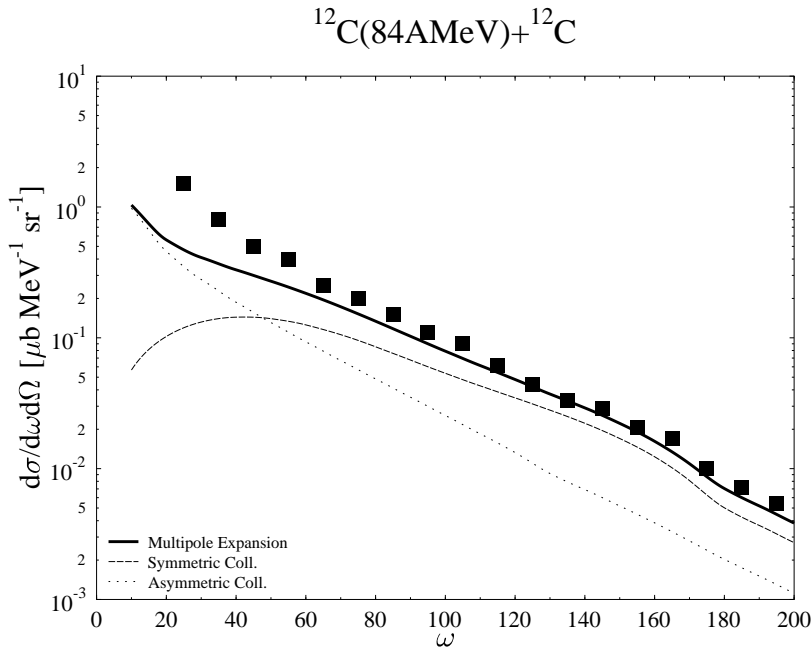,height=10cm}}
\caption{\label{ccospek}Comparision of the measured energy distribution at
$\theta_k=90^\circ$ with
the result of the multipole expansion.}
\end{figure}

One possible signal of the slope of the energy spectrum in Figure
\ref{ccospek} is, that the time of 
the collision is much shorter than one might infer from the naive estimation,
simply regarding the time the nuclei need to pass the distance
$D/\gamma_i$, which gives approximately 25fm/c for the considered process.
Figure \ref{ccospek} shows an arithmetically 
averaged spectrum for collision times
between 6 and 16fm/c. The gap at soft photons could be closed 
by adjusting the ratio of symmetric to asymmetric collisions.\\

According to Fig. \ref{ccospek} and 
the considerations made above, the angular distribution should
change with photon energy. 
To produce a photon with high energy, one
needs many particles which deposit their energy in one photon. Since the
probability to form large clusters depends on the number of involved
particles, collisions with small impact parameters, i.e. more symmetric
collisions, contribute more to the high energy tail of the spectrum. 

For the calculation we used the following values for the parameters: For the
mean total momentum we assumed 320 AMeV. The
final velocity was taken to be 8\% of the initial velocity, $\Delta t$ is
approximately one third of the actual, $b$ dependent collision time. Since
the final velocity is very small, the effects of the transverse scattering
are small as well. The scattering angle was choosen to be $14^\circ$.
These values gave the observed agreement with the data. 

\section{Conclusion}
The multipole expansion of electromagnetic radiation as already presented in
\cite{Bouwkamp} applied to bremsstrahlung provides a transparent
understanding of the radiated fields. Compared to the full calculation 
the isolated handling of the multipole
components and their mutual interference allows 
for a deeper insight in the influence of different parameters and
properties of the current as, e.g., transverse scattering and the presence of
symmetries. 

The fair agreement of the calculation presented in this paper with the data 
is no proof for 
the correctness of the model. It demonstrates, however, that
a completely coherent treatment of bremsstrahlung is able to describe the
data. The "exponential" shape of the energy spectrum, which was
interpreted as being caused by incoherent photon emission from the 
uncorrelated motion of
particles in a thermalized nucleon gas \cite{Niefenecker} can alternatively 
be described
by the coherent photon emission of the strongly correlated motion during the
stopping phase in the collision. Our simple model assumptions have to be
supported by an extended calculation, where a detailed stopping mechanism, as
e.g. in \cite{Stahl}, is assumed and the spatial extension of the colliding
particles is taken into account. 

We conclude, that the measured data, especially the integrated cross
section, 
do neither show information about thermalization, though the angular
distribution was even interpreted as representing isotropic radiation
\cite{Stahl}, nor provide a proof for incoherent photon production. 

\section*{Acknowledgements}
U.~Eichmann would like to thank Dr.J.Reinhardt for advice and helpful discussions.

\end{document}